\documentclass[aps,10pt,pre,twocolumn,superscriptaddress,citesort,floatfix]{revtex4-2}
\usepackage{graphicx}
\usepackage{amsmath, amssymb}
\usepackage{overpic}
\usepackage{amsfonts}
\usepackage{tikz}
\usepackage{enumitem}
\usetikzlibrary{backgrounds,arrows.meta,decorations.pathreplacing,positioning,calc,3d,fit}

\usepackage{tabularx}
\usepackage{adjustbox}
\usepackage{orcidlink}

\usepackage{bm}
\usepackage{lipsum}
\usepackage{hyperref}
\usepackage{cleveref} 
\usepackage{xcolor}
\definecolor{data_blue}{HTML}{1F78B4}
\definecolor{data_red}{HTML}{E31A1C}
\definecolor{data_orange}{HTML}{ff7f0e}
\definecolor{data_green}{HTML}{2ca02c}
\definecolor{tab_blue}{HTML}{1F77B4}
\hypersetup{ breaklinks=true, colorlinks=true, linkcolor=tab_blue, filecolor=tab_blue, urlcolor=tab_blue,
 citecolor=tab_blue,
}
\usepackage{pgfplots}
\pgfplotsset{compat=1.17}

\newcommand{\calc}[1]{\pgfmathparse{#1}\pgfmathprintnumber{\pgfmathresult}}

\usepackage[most]{tcolorbox}
\tcbset{on
  line, boxsep=1pt, left=0pt,right=0pt,top=0pt,bottom=0pt,
  colframe=gray!20, colback=gray!20, highlight math style={enhanced} }
\newcommand{\figlabel}[1]{\tcbox{#1}}

\begin{document}

\title{The multi-fractal nature of pedestrian arrival times}

\author{Caspar A.S. Pouw\orcidlink{0000-0002-3041-4533}}
\email{c.a.s.pouw@tue.nl}
\affiliation{Fluids and Flows group and J.M. Burgers Center for Fluid Mechanics, Eindhoven University of Technology,
 5600 MB Eindhoven, Netherlands}
\affiliation{ProRail BV, Moreelsepark 2, Utrecht, 3511EP, The Netherlands}

\author{Alessandro Corbetta\orcidlink{0000-0001-6979-3414}}
\affiliation{Fluids and Flows group and J.M. Burgers Center for Fluid Mechanics, Eindhoven University of Technology, 5600 MB Eindhoven, Netherlands}

\author{Alessandro Gabbana\orcidlink{0000-0002-8367-6596}}
\affiliation{Department of Physics and Earth Sciences, University of Ferrara, 44122 Ferrara, Italy}
\affiliation{INFN Ferrara, 44122 Ferrara, Italy}

\author{Federico Toschi\orcidlink{0000-0001-5935-2332}}
\affiliation{Fluids and Flows group and J.M. Burgers Center for Fluid Mechanics, Eindhoven University of Technology,
 5600 MB Eindhoven, Netherlands}
\affiliation{CNR-IAC, I-00185 Rome, Italy}

\begin{abstract} 
Pedestrian arrival times exhibit complex temporal organization across multiple scales, shaped by working hours,
transportation schedules, and collective behaviors -- features often neglected in conventional pedestrian arrival models. 
Using a dataset comprising over 23 million pedestrian movements at a Dutch railway station, we show that arrival processes 
cannot be fully characterized by inter-arrival time statistics alone. Instead, we demonstrate that pedestrian arrivals 
exhibit clear multifractal scaling, revealing scale-dependent correlations across a broad range of timescales.

To quantify these properties, we apply a framework based on generalized fractal dimensions, which captures the
heterogeneous structure of arrivals beyond standard point-process descriptions. This approach enables the
identification of distinct temporal regimes associated with external forcing and provides a
quantitative basis for constructing more realistic synthetic arrival processes.
Beyond pedestrian dynamics, this approach offers methodological relevance for understanding non-trivial 
arrival processes in other physical or biological systems. 

\end{abstract}

\maketitle

\section{Introduction}\label{sec:intro}
Many natural and social systems exhibit dynamics that unfold across multiple scales~\cite{wilson1979problems}. 
Two classical examples are given by fluid dynamics turbulence~\cite{benzi-pr-2023}, where velocity fluctuations
and coherent structures span a wide range of scales, from  integral  down to the dissipative scale~\cite{frisch-book-1995}, 
and by the large-scale structure of the Universe in cosmology, where matter distribution shows correlations from 
galaxies to clusters and superclusters~\cite{peebles-book-2020}. 
But is there  physics at every scale? Whether different physical laws govern each observational scale strongly 
depends on the system considered. When spatial or temporal correlations decay rapidly, then no structure emerges 
and the system appears as effectively random. On the other hand, when long-range correlations persist, 
non-trivial statistical patterns arise that demand scale-dependent descriptions, whose analysis requires
multi-scale or even multi-fractal approaches.

Pedestrian dynamics provides a striking example of such multi-scale complexity. In this work, we focus our
attention on the arrival times of passengers in a train station. This process is expected to be intrinsically
heterogeneous in time because, to give a few examples, daily commuting cycles, weekday/weekend differences, seasonal
variations, and holidays clearly shape arrival patterns. 
Consequently, it is evident that the system does not exhibit translational invariance with respect to time, 
nor can it be fully described by simple homogeneous models. Instead, pedestrian arrivals display burstiness, 
clustering, and scaling properties beyond the simple assumptions often used in pedestrian dynamics 
simulators~\cite{karsai-book-2018, liu-trr-2022}.
Understanding these dynamics is important for practical applications. Growing demand for public transportation leads to
increasingly intense and synchronized pedestrian flows, particularly during peak hours~\cite{dutchgov-2019}. Accurate
models of pedestrian arrivals are therefore essential for assessing the performance, safety, and resilience of
transport infrastructures. However, many existing approaches assume Poissonian arrivals, which neglect correlations and
temporal structure observed in real data.

A natural framework for investigating these phenomena is that of point processes (also known as counting processes).
Point processes provide a flexible tool to describe stochastic events in time and space across disciplines such as
physics, astronomy, biology, geology, and ecology~\cite{daley-stat-1998, diggle-book-2013}. The simplest point process, 
the Poisson process, assumes independence between events, implying the absence of correlations. However, many systems exhibit
dependencies: earthquakes and their aftershocks, epidemics, or criminal activity often display clustering that is well
described by self-exciting processes such as the Hawkes process~\cite{hawkes-biometrika-1971, corral-pre-2022, meyer-aas-2014, mohler-jasa-2011}. 
In such systems, the occurrence of one event increases the probability of subsequent nearby events, giving rise 
to larger structures across scales.

Across many complex systems, the marginal inter-arrival time distribution $P(\tau)$ has been employed 
to characterize flow dynamics. For example, in granular and crowd flows passing through bottlenecks, 
Zuriguel and collaborators demonstrated that a transition of $P(\tau)$ from exponential to power-law tails, 
observable across systems as different as granular grains, sheep herds, and pedestrian crowds, 
signals the onset of a jamming phase transition at bottlenecks~\cite{zuriguel-sr-2014, garcimartin-pre-2015,souzy-pre-2020}. 
The setting considered here is complementary, as the flows under study occur at densities values well below 
any jamming threshold. Yet, we find that a richer multi-scale structure persists that the marginal 
distribution alone cannot reveal. This motivates moving beyond $P(\tau)$ toward a framework that captures 
correlations across a hierarchy of timescales.

Such a shift in perspective has proven transformative in other fields. For example, in network traffic analysis, 
Ethernet packet arrivals were long modeled as Poisson processes; multifractal analysis revealed long-range dependence and 
self-similar burstiness, fundamentally changing how networks are engineered~\cite{leland-ton-1994, riedi-ton-1999}. 
Similarly, in financial time series, marginal return distributions miss volatility clustering entirely, which 
multifractal models capture~\cite{calvet-rfs-2002}. The present work pursues an analogous program for pedestrian arrivals.

We analyze a high-statistics dataset of pedestrian movements recorded at a railway station, comprising more than 23 million arrivals. 
We first examine inter-arrival time distributions and highlight their limitations in capturing the full temporal organization of the process. 
To overcome these limitations, we adopt a multi-scale approach based on coarse-graining and generalized fractal dimensions. 
This framework allows us to quantify correlations across different temporal scales and to characterize the arrival process in terms of its scaling properties. 
Our results show that pedestrian arrivals exhibit clear multifractal behavior, indicating the presence of scale-dependent variability that cannot be captured by standard point-process models.

The rest of this work is organized as follows: in~\cref{sec:crowd-flow-measurements} we describe the experimental methods, 
in~\cref{sec:iat} we demonstrate the rich phenomenology of the arrival process through inter-arrival distributions, 
in~\cref{sec:mfa} we present the multi-scale and multifractal analysis, 
and in~\cref{sec:conclusion_arrivals} we discuss conclusions and possible future developments. Finally in the Appendix
section we provide a validation of the multi-scale multifractal framework applied to a synthetic signal, showing 
it correctly reproduce known analytical properties.

\section{Experimental methods}\label{sec:crowd-flow-measurements}
\begin{figure}
  \centering
  \includegraphics[width=\columnwidth, trim=0 160 0 40, clip]{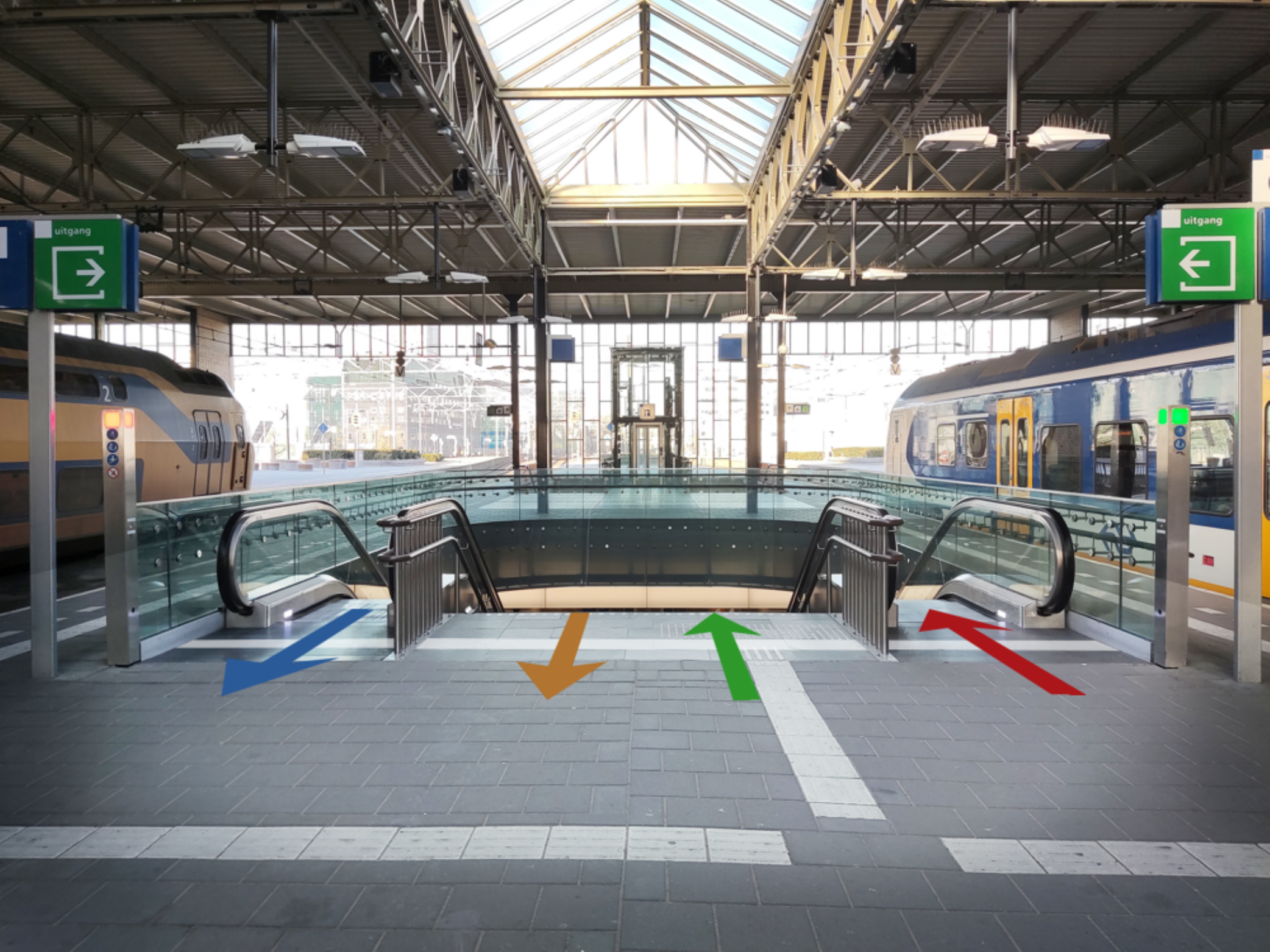}
  \adjustbox{trim = 0 20 0 -10, width=\columnwidth}{
  \begin{tikzpicture}[node distance=0.3cm]
    \node[rectangle] (c1) {};
    \node[rectangle, below=of c1] (c2) {};
    \node[rectangle, below=of c2] (c3) {};
    \node[rectangle, below=of c3] (c4) {};
    \node[rectangle, below=of c4] (c5) {};
    \node[rectangle, below=of c5] (c6) {};

    \node[rectangle, right=3cm of c1] (b1) {};
    \node[rectangle, right=3cm of c2] (b2) {};
    \node[rectangle, right=3cm of c3] (b3) {};
    \node[rectangle, right=3cm of c4] (b4) {};
    \node[rectangle, right=3cm of c5] (b5) {};
    \node[rectangle, right=3cm of c6] (b6) {};
    \node[draw, rectangle, minimum width=3cm, minimum height=1.5cm, rounded corners, thick, black] (platform1) {Train platform};

    \node[left=3cm of platform1] (l1) {};
    \node[above=0.2cm of l1] (l2) {};
    \node[right=3cm of l2] (q1) {};
    \node[below=0.2cm of l1] (l3) {};
    \node[right=3cm of l3] (q2) {};
   
    \draw[{Stealth[data_green]}-, thick, data_green] (l2) to node[fill=white, rounded corners] {Stairs out} (q1);
    \draw[-{Stealth[data_orange]}, thick, data_orange] (l3) to node[fill=white, rounded corners] {Stairs in} (q2);

    \node[right=3cm of platform1] (r1) {};
    \node[above=0.2cm of r1] (r2) {};
    \node[left=3cm of r2] (d1) {};
    \node[below=0.2cm of r1] (r3) {};
    \node[left=3cm of r3] (d2) {};

    \draw[{Stealth[data_blue]}-, thick, data_blue] (d1) to node[fill=white, rounded corners] {Escalator in} (r2);
    \draw[-{Stealth[data_red]}, thick, data_red] (d2) to node[fill=white, rounded corners] {Escalator out} (r3);

    \node[rectangle, below=of platform1] (t3) {};
    \node[rectangle, left=of t3] (t2) {};
    \node[rectangle, left=1.5cm of t2] (t1) {Trains in};
    \node[rectangle, above=of t2] (p1) {};
    \node[rectangle, right=of t3] (t4) {};
    \node[rectangle, right=1.5cm of t4] (t5) {Trains out};
    \node[rectangle, above=of t4] (p2) {};
    
    \draw[-{Stealth[black]}, thick] (t1) -| (p1.south);
    \draw[-{Stealth[black]}, thick] (p2.south) |- (t5);

  \end{tikzpicture}
}
  \adjustbox{trim = 0 0 0 10, width=\columnwidth}{
  \begin{tikzpicture}
   \begin{axis}[
    ybar=0pt,
    bar shift=0pt,
    ymin=0,
    ymax=12500000,
    width=3.54in,
    height=1.5in,
    bar width=40pt,
    ylabel={Arrival count},
    symbolic x coords={
        Stairs in,
        Stairs out,
        Escalator in,
        Escalator out
    },
    xtick={Stairs in, Stairs out, Escalator in, Escalator out},
    enlarge x limits=0.2,
    x tick label style={rotate=20,anchor=east},
    nodes near coords,
    nodes near coords align={vertical},
    ymajorgrids=true,
    grid style=dashed,
    tick label style={/pgf/number format/1000 sep=\,},
]

\addplot[fill=data_orange, draw=data_orange] coordinates {
    (Stairs in,2206256)
};

\addplot[fill=data_green, draw=data_green] coordinates {
    (Stairs out,4789385)
};

\addplot[fill=data_blue, draw=data_blue] coordinates {
    (Escalator in,10122595)
};

\addplot[fill=data_red, draw=data_red] coordinates {
    (Escalator out,6464807)
};

\end{axis}
  \end{tikzpicture}
}
\begin{overpic}[width=\columnwidth]{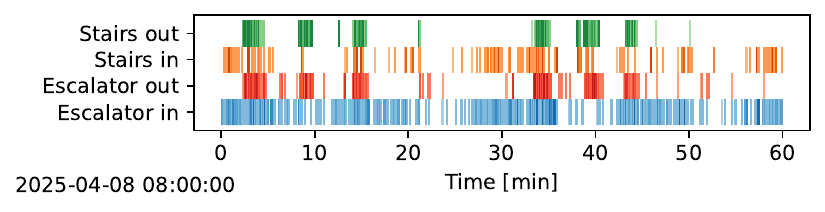}
  \put(0, 148){\figlabel{a}}
  \put(0, 91){\figlabel{b}}
  \put(0, 64){\figlabel{c}}
  \put(0, 22){\figlabel{d}}
  \end{overpic}
  \caption{(a) Eindhoven Centraal Railway Station (the Netherlands), track 3 (left) and track 4
    (right). The main entrance point to the platform features a staircase in the middle with escalators on either side.
    Arrows indicate the direction of pedestrian traffic entering and leaving the platform. (b) The train platform from
    (a) modeled as a single node inside a larger network. Edges model inbound and outbound traffic corresponding to the
    arrows in (a). Inbound traffic ascends, while outbound traffic descends to a lower tunnel. (c) Absolute count of
    pedestrian arrivals between May 1, 2021, and August 1, 2025 differentiated by usage mode and direction. (d) Time
    series representation of the pedestrian traffic over the color coded edge connections for a 60 minute time window.
    Each vertical line represents a pedestrian arrival. Shorter intervals between arrivals have more saturated colors,
    indicating higher traffic density. Clustering in the outbound pedestrian traffic reveals that the arrival process
    is modulated by the train schedule.
    }\label{fig:network} 
\end{figure}
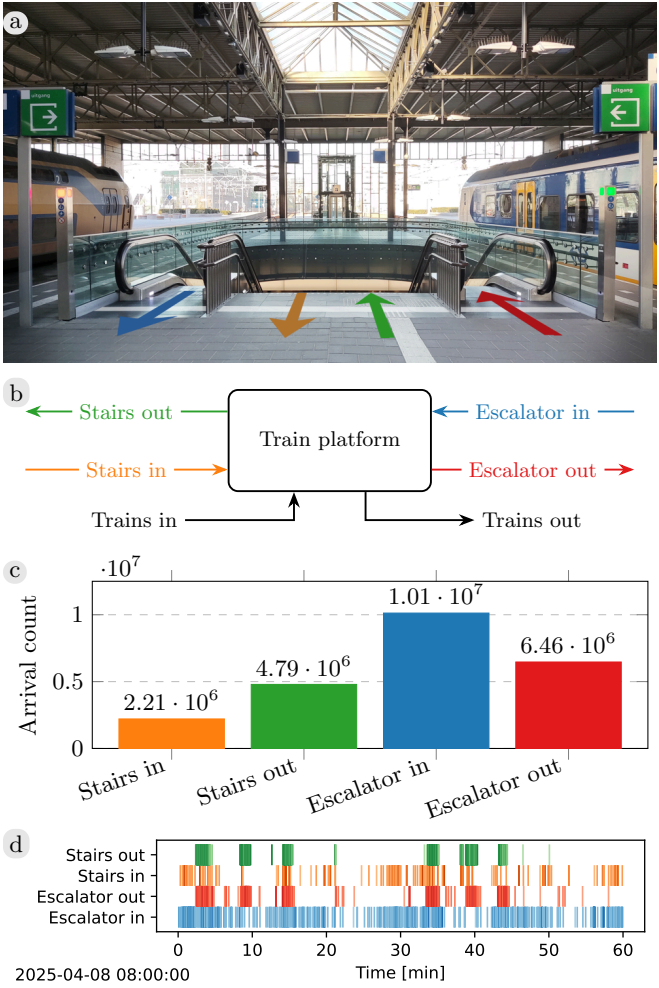

In this work, we use pedestrian measurements on tracks 3 and 4 of Eindhoven Railway Station (the Netherlands)
(cf.~Fig.~\ref{fig:network}(a)) between May 1, 2021, and August 1, 2025. For a detailed overview of the measurement site we refer to Ref.~\cite{pouw-trc-2024}. Data in this work is collected using commercial pedestrian tracking sensors, similar to what is employed in Ref.~\cite{pouw-plos-2020}. A subset of the raw trajectory measurements is publicly available~\cite{pouw-ehvdata-2024}. The integrated line-based flux estimation of the sensors records the time series with line crossing events $t_j = \{t_1, t_2, \dots, t_N\}$ with $t_j$ the time at which pedestrian $j$ steps
over the white line in Fig.~\ref{fig:network}(a).

In Fig.~\ref{fig:network}(b), we model the train platform from Fig.~\ref{fig:network}(a) as a single node with edges
modeling the incoming and outgoing pedestrian traffic flows. The edges match the color of the arrows 
in Fig.~\ref{fig:network}(a). By convention, flows towards the train platform are labeled `in' and flows away from it as `out'.
Note that the staircase connects the platform to a lower tunnel, as such, `in'-flows are ascending, and `out'-flows are
descending.
During the measurement period, we observed a total of more than 23 million pedestrian arrivals, 
see~Fig.~\ref{fig:network}(c) for the count distinguished between usage mode (stairs vs. escalators) 
and direction (inbound vs. outbound). 
In Fig.~\ref{fig:network}(d) we visualize for a 60 minute interval the time series with line crossing measurements; the
events are color coded according to the other figure panels. Each vertical line indicates the moment $t_j$, at which a
pedestrian $j$ crosses the white line. Arrivals with shorter intervals are shown with more saturated colors. Inside a
train station, pedestrian arrivals and departures are naturally influenced by the train schedule, which acts as an
external forcing. The time series in Fig.~\ref{fig:network}(d) reveals clustering in the outbound traffic, indicating
modulation of the arrival process by subsequent train arrivals.

\section{Inter-arrival time}\label{sec:iat} 
\begin{figure}
  \centering
  \begin{overpic}[width=\columnwidth]{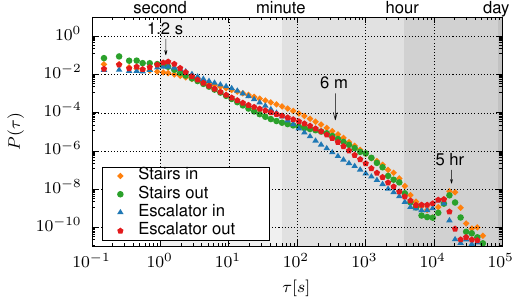}
  \end{overpic}
  \caption{Probability distributions of inter-pedestrian time intervals $\tau$ at Eindhoven Centraal tracks 3 and 4
    for different usage modes and directions. The highlighted local maxima in the curves across multiple temporal scales suggest
    modulations of pedestrian flows at distinct timescales.
   }\label{fig:all-interevent-pdfs}
\end{figure}

To start, we study the arrival sequence using inter-arrival times, defined as the time between two consecutive 
arrivals $\tau_j = t_j - t_{j-1}$~\cite{karsai-book-2018} sometimes described as the waiting time. 
In Fig.~\ref{fig:all-interevent-pdfs} we show probability distributions $P(\tau)$ of inter-arrival times, 
differentiated by usage mode and direction.
The distributions reveal higher relative ``excess'' probabilities for multiple distinct time intervals. For instance, we
observe a local peak at time intervals of $\tau\approx5$ hours, reflecting the time when the train station is closed
overnight (approximately between 00:30 and 05:30). Shorter time intervals, $\tau\approx6$ minutes, align with the
average train arrival frequency (most prominent in the outbound direction), modulating the flow of pedestrians (see also
Fig.~\ref{fig:network}(d)). The time interval observed most frequently is $\tau\approx1.2$ seconds, which can be
related to the frequency of the escalator steps.

\paragraph*{Limitations} The inter-arrival time distribution $P(\tau)$ provides useful information about 
the temporal arrival process and its modulations. However, as a marginal distribution, it characterizes 
each interval $\tau_j = t_j - t_{j-1}$ independently and therefore neglects correlations between successive intervals.
A complete statistical characterization would require the joint distribution $P(\tau_{j-m}, \dots, \tau_{j+m})$, 
which is generally intractable. 
For example, processes with identical marginal $P(\tau)$ but different temporal orderings remain indistinguishable
at this level of description, even though their underlying dynamics may differ substantially. 
Furthermore, long-time-scale correlations between bursts of arrivals can be masked by the accumulation of many small events,
which effectively act as noise in this type of statistical analysis.
To address these limitations, we introduce in the next section a multi-scale framework that characterizes the arrival
process through its scaling properties across a broad range of temporal resolutions.

\begin{figure}
  \resizebox{\columnwidth}{!}{
\begin{tikzpicture}[
  leaf/.style={draw,rectangle,minimum size=7mm,inner sep=2pt},
  shadeval/.code={
    \pgfmathsetmacro{\shade}{int(100*(#1)/15)}
    \edef\tikz@temp{data_blue!\shade}
    \tikzset{fill=\tikz@temp}
    \ifnum\shade>50
      \tikzset{text=white}
    \else
      \tikzset{text=black}
    \fi
  },
  levelbox/.style={draw=none,rectangle,minimum size=7mm,inner sep=2pt}, 
  edge from parent/.style={draw,-}
]
\def\ygap{16mm}

\node (b1)  [leaf, shadeval = 15] {1};
\node (b2)  [leaf, right=1mm of b1, shadeval = 0] {0};
\node (b3)  [leaf, right=1mm of b2, shadeval = 15]  {1};
\node (b4)  [leaf, right=1mm of b3, shadeval = 15]  {1};
\node (b5)  [leaf, right=1mm of b4, shadeval = 0]  {0};
\node (b6)  [leaf, right=1mm of b5, shadeval = 0]  {0};
\node (b7)  [leaf, right=1mm of b6, shadeval = 0]  {0};
\node (b8)  [leaf, right=1mm of b7, shadeval = 15]  {1};
\node (b9)  [leaf, right=1mm of b8, shadeval = 15]  {1};
\node (b10) [leaf, right=1mm of b9, shadeval = 15]  {1};
\node (b11) [leaf, right=1mm of b10, shadeval = 0] {0};
\node (b12) [leaf, right=1mm of b11, shadeval = 0] {0};
\node (b13) [leaf, right=1mm of b12, shadeval = 0] {0};
\node (b14) [leaf, right=1mm of b13, shadeval = 0] {0};
\node (b15) [leaf, right=1mm of b14, shadeval = 15] {1};
\node (b16) [leaf, right=1mm of b15, shadeval = 0] {0};

\node (p1) [leaf, fit=(b1)(b2), inner sep=0pt, shadeval = 7] at ($ (b1)!0.5!(b2)   + (0,\ygap) $) {\calc{1/2}};
\node (p2) [leaf, fit=(b3)(b4),inner sep=0pt, shadeval = 15] at ($ (b3)!0.5!(b4)   + (0,\ygap) $) {\calc{2/2}};
\node (p3) [leaf, fit=(b5)(b6),inner sep=0pt, shadeval = 0] at ($ (b5)!0.5!(b6)   + (0,\ygap) $) {0};
\node (p4) [leaf, fit=(b7)(b8),inner sep=0pt, shadeval = 7] at ($ (b7)!0.5!(b8)   + (0,\ygap) $) {\calc{1/2}};
\node (p5) [leaf, fit=(b9)(b10),inner sep=0pt, shadeval = 15] at ($ (b9)!0.5!(b10)  + (0,\ygap) $) {\calc{2/2}};
\node (p6) [leaf, fit=(b11)(b12),inner sep=0pt, shadeval = 0] at ($ (b11)!0.5!(b12) + (0,\ygap) $) {0};
\node (p7) [leaf, fit=(b13)(b14),inner sep=0pt, shadeval = 0] at ($ (b13)!0.5!(b14) + (0,\ygap) $) {0};
\node (p8) [leaf, fit=(b15)(b16),inner sep=0pt, shadeval = 7] at ($ (b15)!0.5!(b16) + (0,\ygap) $) {\calc{1/2}};

\draw (p1)--(b1)  (p1)--(b2)
      (p2)--(b3)  (p2)--(b4)
      (p3)--(b5)  (p3)--(b6)
      (p4)--(b7)  (p4)--(b8)
      (p5)--(b9)  (p5)--(b10)
      (p6)--(b11) (p6)--(b12)
      (p7)--(b13) (p7)--(b14)
      (p8)--(b15) (p8)--(b16);

\node (gp1) [leaf, fit=(p1)(p2), inner sep=0pt, shadeval = 11] at ($ (p1)!0.5!(p2) + (0,\ygap) $) {\calc{3/4}};
\node (gp2) [leaf, fit=(p3)(p4), inner sep=0pt, shadeval = 3] at ($ (p3)!0.5!(p4) + (0,\ygap) $) {\calc{1/4}};
\node (gp3) [leaf, fit=(p5)(p6), inner sep=0pt, shadeval = 7] at ($ (p5)!0.5!(p6) + (0,\ygap) $) {\calc{2/4}};
\node (gp4) [leaf, fit=(p7)(p8), inner sep=0pt, shadeval = 3] at ($ (p7)!0.5!(p8) + (0,\ygap) $) {\calc{1/4}};

\draw (gp1)--(p1) (gp1)--(p2)
      (gp2)--(p3) (gp2)--(p4)
      (gp3)--(p5) (gp3)--(p6)
      (gp4)--(p7) (gp4)--(p8);

\node (ggp1) [leaf, fit=(gp1)(gp2), inner sep=0pt, shadeval = 7] at ($ (gp1)!0.5!(gp2) + (0,\ygap) $) {\calc{4/8}};
\node (ggp2) [leaf, fit=(gp3)(gp4), inner sep=0pt, shadeval = 5] at ($ (gp3)!0.5!(gp4) + (0,\ygap) $) {\calc{3/8}};

\draw (ggp1)--(gp1) (ggp1)--(gp2)
      (ggp2)--(gp3) (ggp2)--(gp4);

\node (root) [leaf, fit=(ggp1)(ggp2), inner sep=0pt] at ($ (ggp1)!0.5!(ggp2) + (0,\ygap) $) {$N/2^m$};
\draw[dashed] (root)--(ggp1) (root)--(ggp2);

\path let \p1 = (b1.west) in coordinate (leftx) at (\x1-10mm,0);

\node (L0) [levelbox] at (leftx |- b1)   {$k=0$};
\node (L1) [levelbox] at (leftx |- p1)   {$k=1$};
\node (L2) [levelbox] at (leftx |- gp1)  {$k=2$};
\node (L3) [levelbox] at (leftx |- ggp1) {$k=3$};
\node (L4) [levelbox] at (leftx |- root) {$k=m$};

\node (t0) at ($ (leftx |- b1) + (0,-\ygap) $) {Time};
\node (t1) at ($ (leftx |- b1) + (2*\ygap,-\ygap) $) {};
\draw[->] (t0)--(t1);

\draw [decorate,decoration={brace,amplitude=5pt,mirror}] 
  ($(b1.south west) + (0,-0.1)$) -- ($(b1.south east) + (0,-0.1)$)
  node[midway,below=6pt] {$l_{0}$};

\draw [decorate,decoration={brace,amplitude=5pt,mirror}] 
  ($(p1.south west) + (0,-0.1)$) -- ($(p1.south east) + (0,-0.1)$)
  node[midway,below=6pt] {$l_1$};
  
\draw [decorate,decoration={brace,amplitude=5pt,mirror}] 
  ($(gp1.south west) + (0,-0.1)$) -- ($(gp1.south east) + (0,-0.1)$)
  node[midway,below=6pt] {$l_2$};

  \draw [decorate,decoration={brace,amplitude=5pt,mirror}] 
  ($(ggp1.south west) + (0,-0.1)$) -- ($(ggp1.south east) + (0,-0.1)$)
  node[midway,below=6pt] {$l_3$};

\draw [decorate,decoration={amplitude=1pt,mirror}] 
  ($(b9.north west) + (0,0.1)$) -- ($(b9.north east) + (0,0.1)$)
  node[midway,above=1pt] {$\mu_{8,0}$};

\draw [decorate,decoration={amplitude=1pt,mirror}] 
  ($(p5.north west) + (0,0.1)$) -- ($(p5.north east) + (0,0.1)$)
  node[midway,above=1pt] {$\mu_{4,1}$};

\draw [decorate,decoration={amplitude=1pt,mirror}] 
  ($(gp3.north west) + (0,0.1)$) -- ($(gp3.north east) + (0,0.1)$)
  node[midway,above=1pt] {$\mu_{2,2}$};

\draw [decorate,decoration={amplitude=1pt,mirror}] 
  ($(ggp2.north west) + (0,0.1)$) -- ($(ggp2.north east) + (0,0.1)$)
  node[midway,above=1pt] {$\mu_{1,3}$};

\end{tikzpicture}
  }
  \caption{Illustration of the coarse-graining renormalization flow for a synthetic signal. The
   leaves of the tree, at $k=0$, illustrate the discrete arrival time series consisting of non-overlapping bins, each
   encoding the arrival of at most a single pedestrian ($1$) or no arrival ($0$) in a short time interval $l_0$. By
   moving up the tree $(k>0)$, the sequence is coarse-grained by increasing the size of the time interval in multiples
   of 2. Each node is inscribed by $\mu_{i,k} = n_{i,k}/2^k$ representing the distribution of arrivals on different
   temporal scales. The process repeats $m$ times until the complete time series is enclosed in a single time interval
   $\mu_{0,m} = N/2^m$.
  }\label{fig:renorm-flow}
\end{figure}
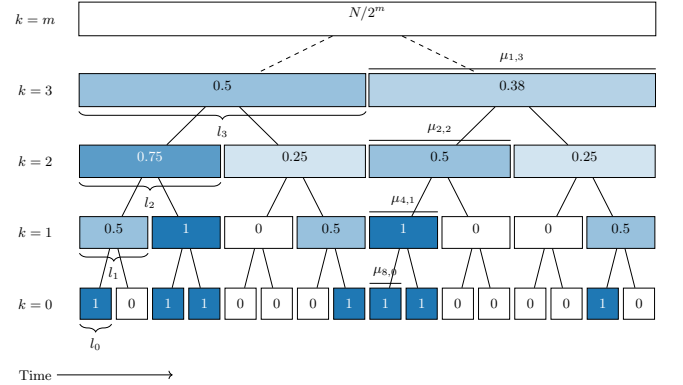

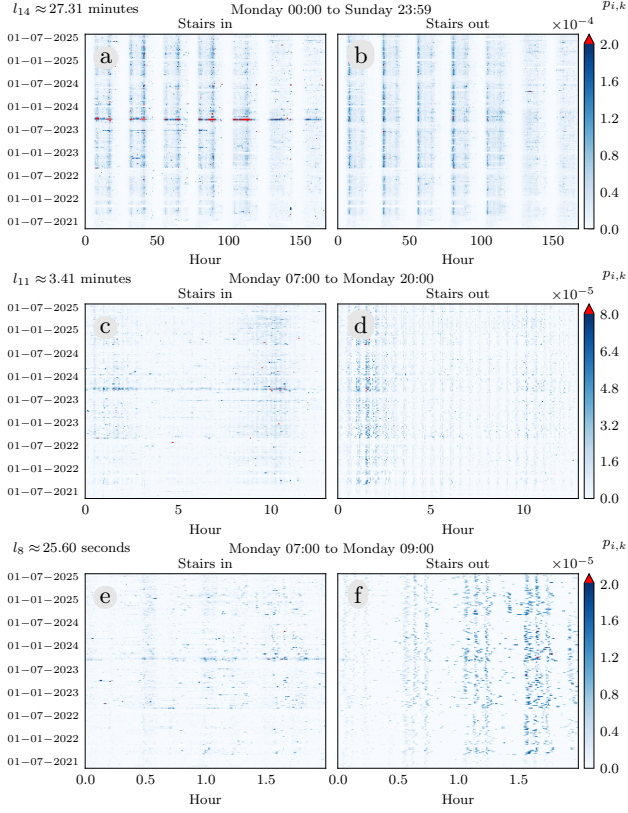
\begin{figure}[tbp]
  \centering \begin{overpic}[width=.95\linewidth]{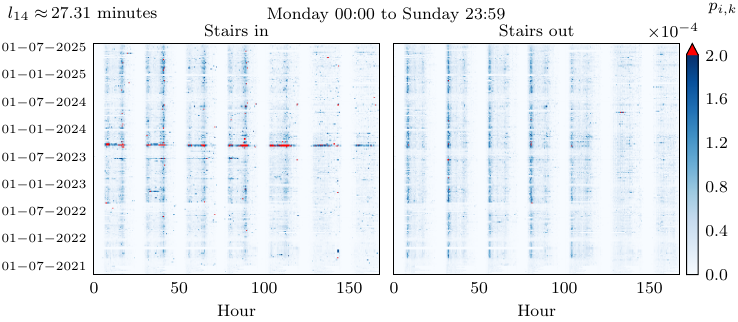}
    \put(14, 33){\figlabel{a}}
    \put(55, 33){\figlabel{b}}
  \end{overpic} \begin{overpic}[width=.95\linewidth]{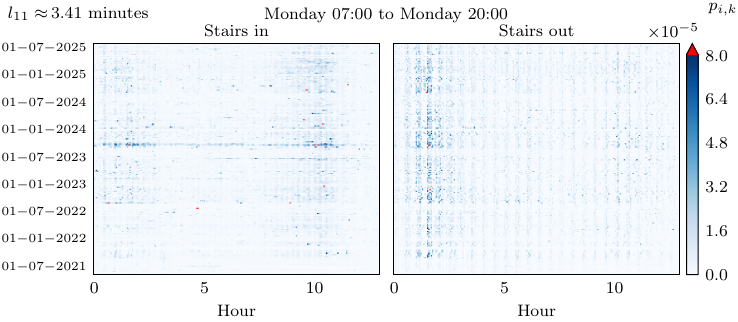}
    \put(14, 33){\figlabel{c}}
    \put(55, 33){\figlabel{d}}
  \end{overpic} \begin{overpic}[width=.95\linewidth]{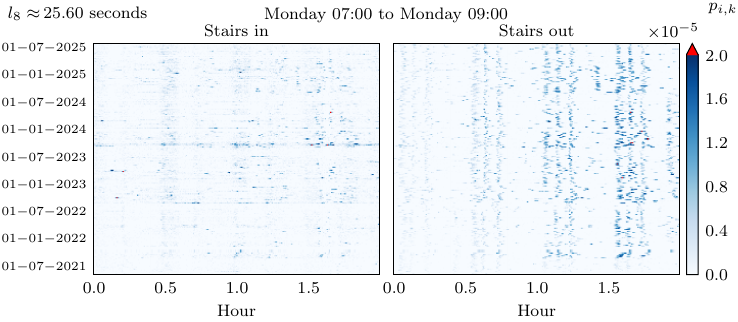}
    \put(14, 33){\figlabel{e}}
    \put(55, 33){\figlabel{f}}
  \end{overpic}
  \caption{ Heatmap of pedestrian traffic on the staircase of Eindhoven Centraal,
   left and right panels show inbound and outbound traffic, respectively. Each bin represents the probability 
   $p_{i,k}=n_{i,k}/N$ per time interval $l_k$ (cf. Eq.~\ref{eq:arrival_prob}). Darker blue indicates higher probability.
   Values outside the color range are shown in red. (a, b) Time intervals of $l_k \approx 27$ minutes, (c, d) 
   time intervals of $l_k \approx 3.5$ minutes, and (e, f) time intervals of $l_k \approx 26$ seconds. 
   We observe regular periodic variations, highlighting the non-uniform distribution across scales. 
   }\label{fig:discretized_arrival_process}
\end{figure}

\section{Multi-scale analysis}\label{sec:mfa} 
To overcome the limitations of inter-event analysis, we adopt a multi-scale framework~\cite{telesca-csf-2004, salat-physa-2017, corral-pre-2022} based on the construction of a scale-dependent measure of arrival density. This is coupled with a multifractal analysis to quantify the statistical properties of the process across temporal scales.

\paragraph{Coarse-grain renormalization} We begin by discretizing the arrival time series $t_j = \{t_1, t_2, \ldots, t_N\}$ into $M_0$ non-overlapping time windows of size $l_0$. For each interval $i$, we define $n_i$ as the number of arrivals within that interval, such that the total number of arrivals is $N = \sum_i n_i$. 

We then apply a coarse-graining procedure that increases the interval size by factors of two,
\begin{equation}
  l_{k+1} = l_{k} \cdot 2 \quad \Rightarrow \quad l_k=l_0 \cdot 2^k,
\end{equation}
where $k$ denotes the coarse-graining level. At each iteration, the number of intervals is reduced accordingly as 
$M_{k+1} = M_k/2$. Fig.~\ref{fig:renorm-flow} illustrates this renormalization procedure for a synthetic sequence. 

At scale $k$, we define the coarse-grained arrival density as $\mu_{i,k} = n_{i,k}/2^k$ for interval $i$ of size $l_k$.
This construction provides a multi-scale representation of the arrival process: small values of $k$ capture short-time
fluctuations, while large values of $k$ probe longer-term dynamics.

To characterize the distribution of arrivals across scales, we consider the normalized measure

\begin{equation}\label{eq:arrival_prob}
  p_{i,k} = \frac{n_{i,k}}{N},
\end{equation}
which represents the probability of observing an arrival in interval $i$ at scale $k$. 
In Fig.~\ref{fig:discretized_arrival_process}, we visualize $p_{i,k}$ for different coarse-graining levels. We set the
smallest interval to $l_0 = 100$~ms. At this resolution, a few time windows contain more than one arrival, but smaller
intervals become impractical due to the large size of our dataset. Fig.~\ref{fig:discretized_arrival_process}(a,b) with
intervals of $l_k \approx 27$ minutes reveals weekly arrival patterns, Fig.~\ref{fig:discretized_arrival_process}
(c,d) with intervals of $l_k \approx 3.5$ minutes captures a daily cycle, and Fig.~\ref{fig:discretized_arrival_process}
(e,f) with intervals of $l_k \approx 26$ seconds shows the dynamics of hourly arrivals. The left and right panels
correspond to inbound and outbound traffic, respectively, with darker colors indicating higher arrival probabilities.

\begin{figure}
  \centering \begin{overpic}[width=\linewidth, trim=0 150 0 0, clip]{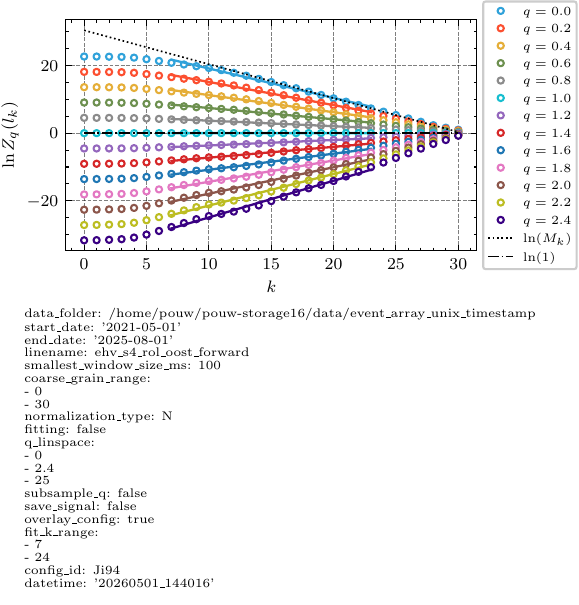}
  \end{overpic}
  \caption{ Partition sums $Z_q(l_k)$ for the pedestrian arrival time series (outbound direction), 
            shown for increasing interval sizes $l_k$ and moment orders $q \in \{0, 0.2, \dots, 2.4\}$. 
            The scaling exponents $\tau(q)$ are obtained from the slopes of linear regressions (solid lines) over the range $6 < k \leq 23$, 
            from which the generalized fractal dimensions $D(q)$ are computed.
          }\label{fig:fit_tau}
\end{figure}

\paragraph{Multi-fractal analysis} Using the probability measure $p_{i,k}$, we compute the partition function for the $q$-th order moments

\begin{equation}
  Z_q(l_k) = \sum_{i} (p_{i,k})^q.
\end{equation} 
Two limiting cases are noteworthy: 1) for $q=1$, all probabilities are summed such that $Z(q=1)=\sum{(p_{i,k})^q}=1$, 
while 2) for $q=0$, the number of intervals of size $l_k$ are counted such that $Z(q=0) = M_k$.

The generalized fractal dimension $D(q)$ is obtained from the scaling exponent $\tau(q)$ as
\begin{equation}\label{eq:generalized_fractal_dimension}
  D(q) = \frac{1}{q-1} \lim_{l \to 0}\frac{\log \sum (p_{i,k})^q}{\log l} = \frac{\tau(q)}{q-1},
\end{equation}
where $\tau(q)$ is defined as
\begin{equation}
  \tau(q) = \lim_{k \to 0}\frac{\ln Z_q}{\ln 2^k}.
\end{equation}
In practice, $\tau(q)$ is estimated by linear regression of $\ln Z_q(l_k)$ as a function of $k$ 
for each value of $q$ (Fig.~\ref{fig:fit_tau}). 
We restrict the fit to the range $6 < k \leq 23$, corresponding to time intervals from $l_7 \approx 13$~\text{s} to $l_{23} \approx 10$~\text{days}.
The lower bound $k>6$ excludes sub-minute scales where discretization artifacts from the $100~\text{ms}$ bin size and occasional
multi-pedestrian crossings within a single bin can bias the scaling. 
The upper bound $k\leq23$ excludes intervals of $l_{24} \approx 10$~days or more, at which point the regular weekly opening-and-closing cycle of the station dominates, reducing the process to a simple two-state modulation and suppressing the multi-scale structure visible at shorter scales.
Within this range, the system operates well below any jamming threshold~\cite{zuriguel-sr-2014}, so interaction-driven
correlations are negligible. The observed structure instead reflects external forcing, such as train schedules and
daily cycles. The multifractal analysis therefore focuses on intermediate to long timescales 
($13~\text{s} \mathrel{\scriptstyle \lesssim} l_k \mathrel{\scriptstyle \lesssim} 10~\text{days}$), where this externally driven temporal organization is most pronounced.

\begin{figure}[t]
  \centering 
  \begin{overpic}[width=\columnwidth, trim=0 129 0 0, clip]{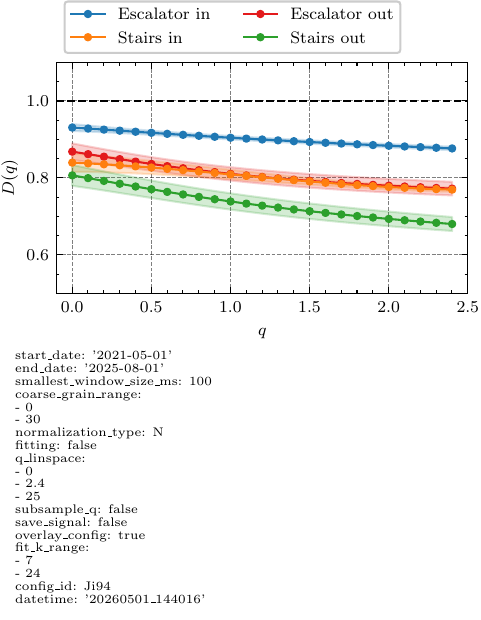}
  \end{overpic}
  \caption{ The generalized fractal dimension $D(q)$ as a function of $q$ derived using 
   Eq.~\ref{eq:generalized_fractal_dimension} differentiated by usage mode and direction. The scaling exponent $\tau(q)$ 
   is obtained from the slope of the linear regression of the partition sum (cf. Fig.~\ref{fig:fit_tau}). Shaded regions
   indicate the 68\% confidence interval derived from the fitting error. For an uncorrelated Poisson distributed
   process the generalized fractal dimension satisfies $D(q)=1$ indicated with a black dashed line. 
   }\label{fig:generalized_fractal_dimension}
\end{figure}

The function $D(q)$ characterizes how arrival probabilities are distributed across scales. Higher values of $q$ weight 
higher densities of arriving pedestrians, while lower values of $q$ probe sparse or low-activity regions.
Fig.~\ref{fig:generalized_fractal_dimension} shows the resulting fractal spectra. A constant $D(q)$ indicates monofractal behavior, whereas a dependence on $q$ signals multifractality~\cite{diego-ras-1999}. 
For an uncorrelated Poisson process, one expects $D(q)=1$ for all $q$. 

Our results clearly deviate from this reference, demonstrating that pedestrian arrivals cannot be described by a single scaling exponent. Instead, multiple exponents are required, confirming the multifractal nature of the process. 
For increasing $q$ (i.e. dense events), staircase flows display stronger intermittency than escalator flows, consistent with the higher usage of escalators. 
Finally, we observe systematic differences between flow directions. Inbound flows exhibit values of $D(q)$ closer to unity, indicating a more homogeneous distribution of arrivals, whereas outbound flows show stronger deviations, consistent with enhanced temporal structuring. This difference likely reflects the modulations induced by the train arrivals, which generate short periods of high pedestrian density.

\section{Conclusion}\label{sec:conclusion_arrivals}

In this study, we conducted a comprehensive multi-scale and multi-fractal analysis of the arrival times of
individuals at a train station, using a dataset covering more than four years and over 23 million pedestrian crossings. 
By calculating the generalized fractal dimension, we demonstrated that the system's behavior cannot be adequately
described by a simple Poisson process, which is often assumed in pedestrian simulators. 

This insight has significant implications for the field of pedestrian simulation. Many algorithms rely on simplified
assumptions, such as the Poisson process, to generate boundary conditions for modeling pedestrian arrivals. Our
findings show that such simplifications do not capture the realistic dynamics of human arrivals in real-world
environments with possible important consequences for the reliability of the simulated scenarios. By incorporating the
generalized fractal dimension into the modeling process, one can potentially better test the quality and realism of
algorithms for generating boundary conditions, providing a more reliable framework for simulating pedestrian behavior.

Moreover, our results open up new avenues for generating data that more accurately reflects real-world patterns. Similar
to how a multiaffine process is constructed to model turbulence~\cite{benzi1993random,biferale1998mimicking,benzi2003intermittency}, 
it is possible to generate synthetic arrival sequences that respect the generalized fractal dimension, providing more accurate input for simulations.
This could lead to more robust and realistic pedestrian models, offering better predictions and more effective designs 
for spaces where crowd movement is a critical factor.

Beyond pedestrian dynamics, the framework developed here is applicable to any point process exhibiting non-trivial
temporal correlations. Natural candidates include seismic sequences~\cite{telesca-csf-2004}, where aftershock
clustering produces multi-scale behavior analogous to train-induced pedestrian bursts, as well as neural spike trains
and vehicular traffic flows, where external rhythms similarly impose structured intermittency on an otherwise
stochastic baseline.

\medskip

\section*{Acknowledgements}
This work is part of the HTSM research program “HTCrowd: a high-tech
platform for human crowd flows monitoring, modeling and nudging” with
project number 17962, and the VENI-AES research program “Understanding
and controlling the flow of human crowds” with project number 16771,
both financed by the Dutch Research Council (NWO).

\section*{Data availability}
A subset of the raw trajectory data is publicly available on
Zenodo~\cite{pouw-ehvdata-2024}. The remaining data cannot be fully published online due to
data ownership restrictions and file size limitations, but may be made
available to qualified researchers upon reasonable request to the
corresponding author, subject to the data provider’s terms and
conditions.

\section*{Ethics statement}
This study has been approved by the Ethical Review Board of Eindhoven
University of Technology (ref. ERB2020AP1, 21st, Feb. 2020). During
this study, pedestrian trajectory data were collected using
anonymous-by-design commercial sensors. The dataset considered
contains only individual trajectories, with no additional personal
features available and/or stored.

\medskip
\appendix

\section{Validation: analytic result for a binomial cascade}\label{app:validation}
To validate the computation of the generalized fractal dimensions, we apply our algorithm to a synthetic signal with a
known, analytically derived fractal dimension. We generate this synthetic signal using a multiplicative
cascade, specifically a binomial cascade, which is essentially the time-reversal of the coarse-graining procedure 
discussed in~\cref{sec:mfa}.

\subsection{Signal construction}

We start with a single interval of length $L$ that contains $N$ arrivals. 
The interval is then split into two equally sized sub-intervals, and the $N$ arrivals are distributed between 
them according to a binomial distribution with probability $p \in (0,1)$:
\begin{align}
  N_{\text{left}} \sim \mathrm{Binomial}(N, p) \\
  N_{\text{right}} = N - N_{\text{left}}
\end{align}
so that the total count is exactly preserved at each split. This procedure is repeated recursively for $\kappa$ levels,
yielding $2^\kappa$ bins, each of length $l_\kappa = L/2^\kappa$, and each containing $n_{i,\kappa}$ arrivals.

The parameter $p$ controls the degree of asymmetry: for $p=0.5$ the cascade is symmetric and, on average, arrivals are
distributed uniformly; for $p \neq 0.5$ the cascade is asymmetric and produces a heterogeneous, intermittent signal
whose multifractal properties are controlled by $p$.

\begin{figure}[t]
  \centering 
  \begin{overpic}[width=\columnwidth, trim=0 50 0 0, clip]{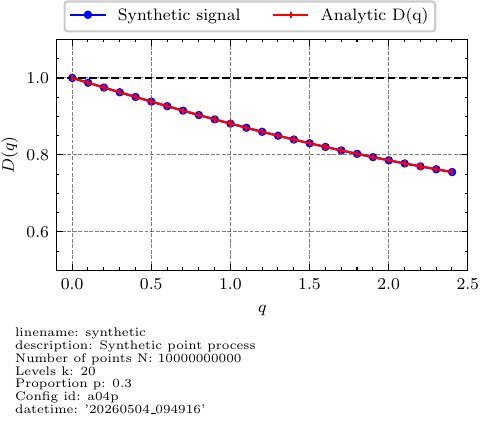}
  \end{overpic}
  \caption{ Generalized fractal dimension $D(q)$ of a synthetic binomial cascade signal with $N=10^{10}$ 
            arrivals, $\kappa=20$ levels, and splitting probability $p=0.3$. The analytic prediction 
            of \eqref{eq:Dq_analytic} is shown alongside the numerical estimate, demonstrating excellent 
            agreement across the full range of $q$.
          }\label{fig:synthetic}
\end{figure}

\medskip

\subsection{Analytic derivation of $\tau(q)$ and $D(q)$}

To derive the expected fractal dimension we compute the partition sum $Z_q$ at cascade level $\kappa$. At each level,
the probability of a given interval depends only on how many times it follows each branch. 
An interval that takes the left branch $j$ times (and the right branch $\kappa-j$ times) has probability
\begin{equation} 
  p_{i,\kappa} = p^j (1-p)^{\kappa - j}.
\end{equation} 
The number of intervals with exactly $j$ left-branches is $\binom{\kappa}{j}$, and therefore the expected partition sum is
\begin{align}\label{eq:Zq_cascade}
  \mathbb{E}\bigl[Z_q(l_\kappa)\bigr]
    &= \sum_{j=0}^{\kappa} \binom{\kappa}{j} \bigl(p^j (1-p)^{\kappa-j}\bigr)^q = \bigl(p^q + (1-p)^q\bigr)^\kappa, 
\end{align} 
Since at level $\kappa$ each bin has size
$l_\kappa = L / 2^\kappa$, we have $\kappa = \log_2(L/l_\kappa)$, and taking the logarithm
of \eqref{eq:Zq_cascade} gives
\begin{equation}
    \ln \mathbb{E}\bigl[Z_q\bigr] = \kappa \ln\bigl(p^q + (1-p)^q\bigr).
\end{equation} 
Comparing with the definition $\tau(q) = \ln Z_q / \ln(L/l_\kappa) = \ln Z_q / (\kappa \ln 2)$,
we read off directly
\begin{equation}\label{eq:tau_analytic}
  \tau(q) = \frac{\ln\bigl(p^q + (1 - p)^q\bigr)}{\ln 2} = \log_2 \bigl(p^q + (1-p)^q\bigr).
\end{equation} 
Substituting into Eq.~\ref{eq:generalized_fractal_dimension} yields the generalized fractal dimension
\begin{equation}\label{eq:Dq_analytic} 
  D(q) = \frac{\tau(q)}{q-1} = -\frac{\log_2 \bigl(p^q + (1 - p)^q\bigr)}{q - 1}.
\end{equation} 
Note that for $p=0.5$, $p^q+(1-p)^q = 2 \cdot 2^{-q}$, so $\log_2(p^q+(1-p)^q) = 1-q$ and $D(q) = 1$ for all $q$,
showing that the uniform cascade is monofractal with dimension one, as expected for a homogeneous process. 
Moreover, in the limit $q\to1$, the above reduces to 
$D(1) = -[p\log_2 p + (1-p)\log_2(1-p)]$, i.e.\ the Shannon entropy of the splitting probability.

\subsection{Numerical validation}

In Fig.~\ref{fig:synthetic} we compare the fractal dimension computed by our algorithm against the analytic prediction
of \eqref{eq:Dq_analytic} for a cascade with $N=10^{10}$ arrivals, $\kappa=20$ levels, and $p=0.3$. The numerical and
analytic curves are in excellent agreement across the full range of $q$, confirming that our implementation correctly
recovers the generalized fractal dimension spectrum.

\bibliography{biblio}

\end{document}